\documentclass[twocolumn,
ams,showkeys,
preprintnumbers,amsmath,amssymb]{revtex4}

\usepackage{graphicx}
\usepackage{dcolumn}
\usepackage{bm}

\usepackage{graphicx}
\usepackage{epsfig}
\usepackage{color}

\begin{document}

\title{Nongauge bright soliton of the nonlinear Schr\"odinger (NLS) equation and a family of generalized NLS equations}

\author{M. A. Reyes and D. Guti\'errez-Ruiz}
\affiliation{Departamento de F\'{\i}sica, Universidad de Guanajuato, DCI Campus Le\'on, Apdo. Postal E143, 37150 Le\'on, Gto., M\'exico}

\author{S. C. Mancas}

\affiliation{Department of Mathematics, Embry-Riddle Aeronautical University,\\ Daytona Beach, FL. 32114-3900, U.S.A.}

\author{H. C. Rosu}
\email{hcr@ipicyt.edu.mx}
\affiliation{IPICyT, Instituto Potosino de Investigacion Cientifica y Tecnologica,\\
Camino a la presa San Jos\'e 2055, Col. Lomas 4a Secci\'on, 78216 San Luis Potos\'{\i}, S.L.P., Mexico}

\begin{abstract}
We present an approach to the bright soliton solution of the NLS equation from the standpoint of introducing a constant potential term in the equation.
We discuss a `nongauge' bright soliton for which both the envelope and the phase depend only on the traveling variable.
We also construct a family of generalized NLS equations with solitonic ${\rm sech}^p$ solutions in the traveling variable and find an exact equivalence with other nonlinear equations, such as the Korteveg-de Vries and Benjamin-Bona-Mahony equations when $p=2$.\\
\end{abstract}
%
%


\hspace{5cm}  Mod. Phys. Lett. A 31 (2016) 1650020

\maketitle



\noindent {\bf 1. Introduction}\\

 For about fifty years now, the one-dimensional cubic NLS equation has been overwhelmingly used for the description of wave propagation in nonlinear optics, fluids, and plasmas \cite{nlse-poe} and arguably it is the best studied nonlinear equation. On the other hand, linearly-extended forms of NLS equations are used by the majority of authors to describe the Bose-Einstein condensates \cite{gross}. Perhaps the simplest form of the latter class is \cite{focus}
\begin{equation}
 i \hbar\frac{dq}{dt}= - \frac{\hbar^2}{2m}\frac{d^2q}{dx^2} - \gamma \left( |q|^{2}-q_0^{2} \right) q~,
\label{efocus}
\end{equation}
where $q_0=$ constant, $\gamma >0$ is the nonlinear strength parameter, and $\hbar$ and $m$ are the Planck constant divided by 2$\pi$ and the mass of the condensate, respectively.
The standard NLS equation is obtained from (\ref{efocus}) through the trivial change of dependent variable (gauge)
\begin{equation}\label{gauge}
q(x,t)=e^{i\gamma q_0^2 t/\hbar} \, \psi(x,t)~.
\end{equation}
In this work, we have two goals. The first one is concerned with the {\em localized} solutions of standard and linearly-extended NLS equations. These localized solutions, of which solitons are a subset, have zero boundary conditions as $x\to\pm\infty$. The NLS soliton solutions are of course well known but we will point here to what we think to be a rather ignored feature of these solutions, namely that there exist a bright soliton of equations of type (\ref{efocus}) which cannot be obtained from the standard bright soliton by the gauging (\ref{gauge}).
Secondly, we construct a family of generalized NLS equations with soliton solutions and find that it is possible to obtain an exact equivalence of these equations with the Korteveg-deVries (KdV) and Benjamin-Bona-Mahony (BBM) equations.

\bigskip

\noindent {\bf 2. The nongauge bright soliton}\\

\begin{figure} [htpb]
\begin{center}
\includegraphics[width=0.35\textwidth]{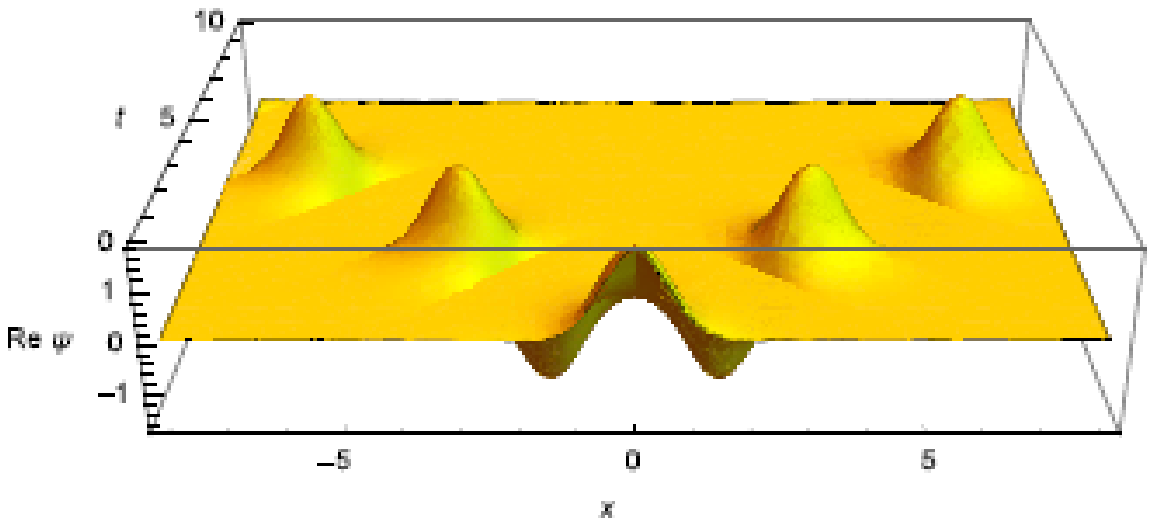} \ \
\includegraphics[width=0.35\textwidth]{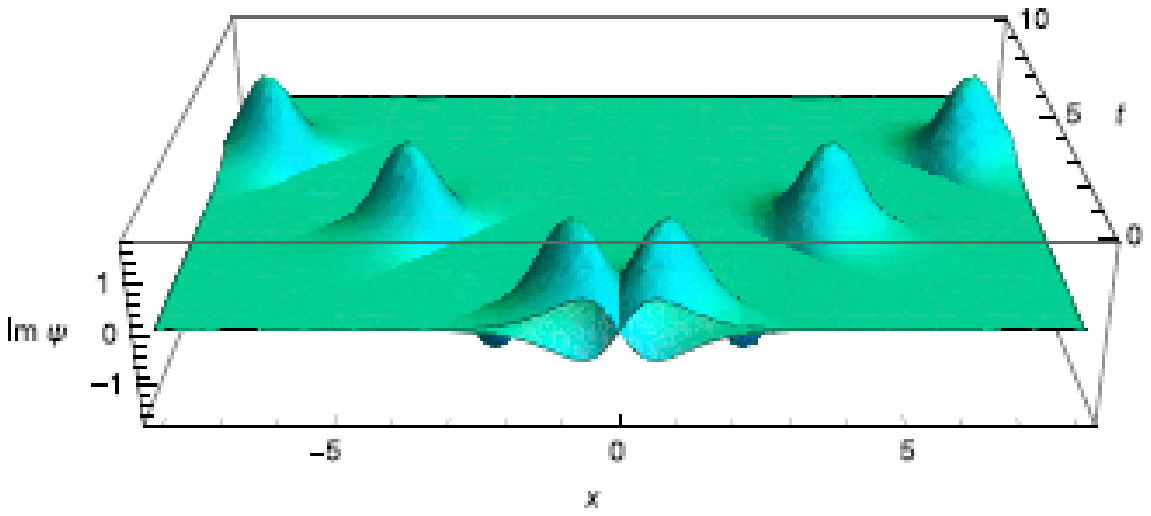}
\includegraphics[width=0.35\textwidth]{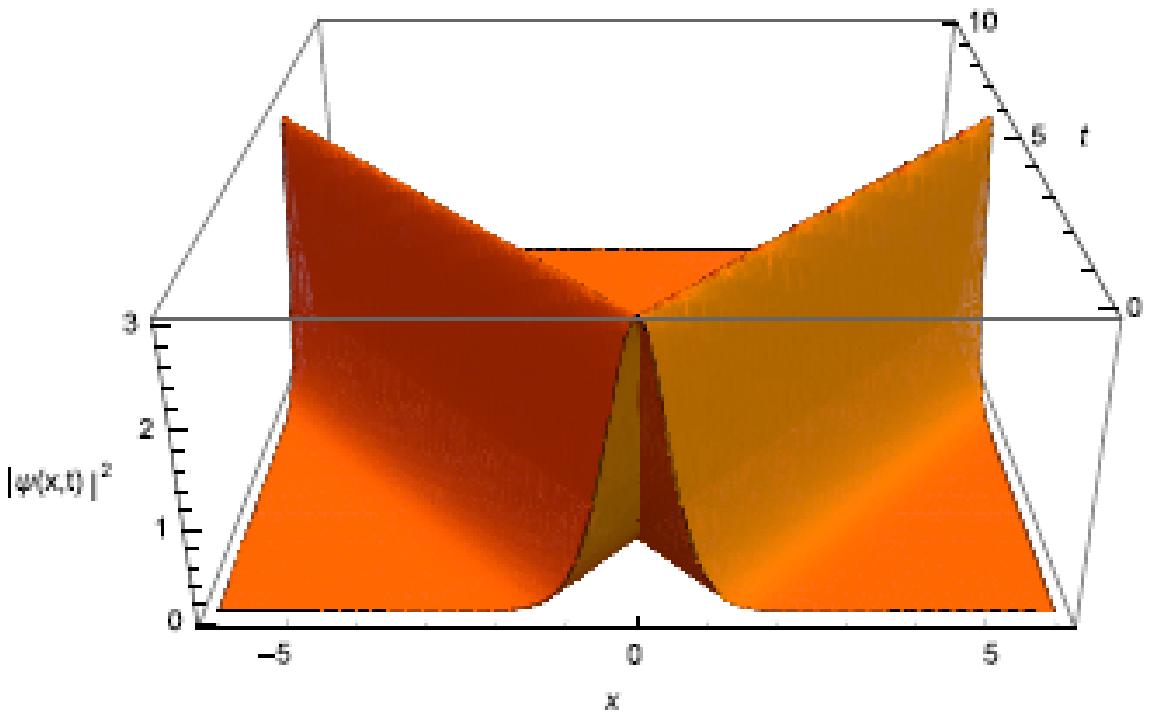}
\caption{\label{stdw1}
(Color online) Plots of the oscillatory real and imaginary parts of the wave function for $\alpha=2$, 
for the case of dimensionless variables, with $z_{0}=\phi_{0}=0$, and $\hbar=\gamma=m=1, v=\mp1$ ,
and its squared modulus. For $v=-1$, the wave is traveling to the left, and for $v=1$ to the right.}
\end{center}
\end{figure}

By applying (\ref{gauge}) to (\ref{efocus}), we obtain the standard NLS equation with no potential
\begin{equation}\label{eq:enls}
i\hbar\frac{\partial\psi}{\partial t}=-\frac{\hbar^{2}}{2m}\frac{\partial{}^{2}\psi}{\partial x^{2}} -\gamma|\psi|^{2}\psi~,
\end{equation}
%
The localized solution of this standard NLS equation is given by
\begin{equation}\label{sol-std}
\psi(x,t)=\sqrt{\frac{2{\cal A}}{\gamma}}
\,\mathrm{sech}\!\left[\sqrt{\frac{2m{\cal A}}{\hbar^{2}}}\,\,
(z+z_{0})\right]
e^{i\left[\frac{mv}{\hbar}z+\phi_{0}+\alpha t\right]}~.
\end{equation}
This solution has been derived by Remoissenet in a different context and notation \cite{Remoissenet}, see also below, and can be obtained as a particular case from the solution given by Katyshev {\em et al} \cite{K78} for NLS equations with arbitrary power nonlinearities of the type $|\psi|^\nu \psi$, $\nu>1$.
In (\ref{sol-std}), ${\cal A}$ is an energy-like quantity given by
$$
{\cal A}=\hbar\alpha-\frac{mv^2}{2}~,
$$
$z=x-vt$ is the traveling variable with speed $v$, $z_{0}$ and $\phi_{0}$ are real integration constants, while $\alpha$ is an angular frequency parameter used to obtain the localized solution. Since $\gamma>0$, and for ${\cal A}>0$, (\ref{sol-std}) can be identified with a bright soliton solution \cite{bright}.
Soliton (\ref{sol-std}) is plotted in Fig.~\ref{stdw1} for the values of the parameters given in the caption, where the oscillatory real and imaginary parts are displayed as well as its squared amplitude.

\medskip

By including a constant potential $V_0$, equation~(\ref{eq:enls}) changes to
\begin{equation}\label{eq:enls2}
i\hbar\frac{\partial\psi}{\partial t}=-\frac{\hbar^{2}}{2m}\frac{\partial{}^{2}\psi}{\partial x^{2}}
-\gamma|\psi|^{2}\psi +V_0\psi~,
\end{equation}
%
whose solution can be obtained by using the ansatz
\begin{equation}\label{trvw}
\psi(x,t)=A(z)e^{i[\phi(z)+\alpha t]}~,
\end{equation}
where $A$ and $\phi$ are functions only of the traveling variable $z$. Using (\ref{trvw}) in (\ref{eq:enls2}) and separating the real and imaginary parts we obtain the system of equations
\begin{equation}\label{eq:catn}
\frac{\mathrm{d^{2}}A}{\mathrm{d}z^{2}}-A\left(\frac{\mathrm{d}\phi}{\mathrm{d}z}\right)^{2}-
\frac{2m(\hbar \alpha+V_{0})}{\hbar^{2}}A+\frac{2m\gamma}{\hbar^{2}}A^{3}+\frac{2mv}{\hbar}A\frac{\mathrm{d}\phi}{\mathrm{d}z}=0~,
\end{equation}
\begin{equation}
A\frac{\mathrm{d^{2}}\phi}{\mathrm{d}z^{2}}+2\frac{\mathrm{d}A}{\mathrm{d}z}\left(\frac{\mathrm{d}\phi}{\mathrm{d}z}-\frac{mv}{\hbar}\right)=0~.
\label{eq:quin}
\end{equation}
Employing a linear phase
$\phi(z)=\frac{mv}{\hbar}z+\phi_{0}$, which satisfies (\ref{eq:quin}) identically,
the  amplitude equation (\ref{eq:catn}) becomes
\begin{equation}
\frac{\mathrm{d^{2}}A}{\mathrm{d}z^{2}}-
\frac{2m{\cal B}}{\hbar^2}A+\frac{2m\gamma}{\hbar^{2}}A^{3}=0~.\label{eq:compr}
\end{equation}
This elliptic equation is easily integrated by assuming zero boundary conditions to obtain  
\begin{equation}\label{eq:solucionfinal}
\psi(x,t)=\sqrt{\frac{2{\cal B}}{\gamma}} 
\,\mathrm{sech} \! \left[\sqrt{\frac{2m{\cal B}}{\hbar^{2}}}
\left(z+z_0\right)\right]
e^{i\left[\frac{mv}{\hbar}z+\phi_0+\alpha t\right]}~,
\end{equation}
where
%
$$
{\cal B}=\hbar \alpha+\Delta Q~, \quad \Delta Q  = V_{0}-\frac{1}{2}mv^{2}~.
$$
From (\ref{eq:solucionfinal}), one can see that setting $V_0=0$ leads to solution (\ref{sol-std}).

We stress that traveling soliton wave solutions without phase factors that depend only on $x$ or $t$, cannot occur for (\ref{eq:enls}), while in the constant potential case this is possible. Indeed, one can see that setting $\alpha=0$ in the solution (\ref{sol-std}) destroys the solitonic feature of the amplitude by turning it into a secant function, while this is possible in the case of the solution (\ref{eq:solucionfinal}) because the potential $V_0$ maintains the solitonic profile of the amplitude.

\bigskip


\begin{figure} [ht]
\begin{center}
\includegraphics[width=0.49\textwidth]{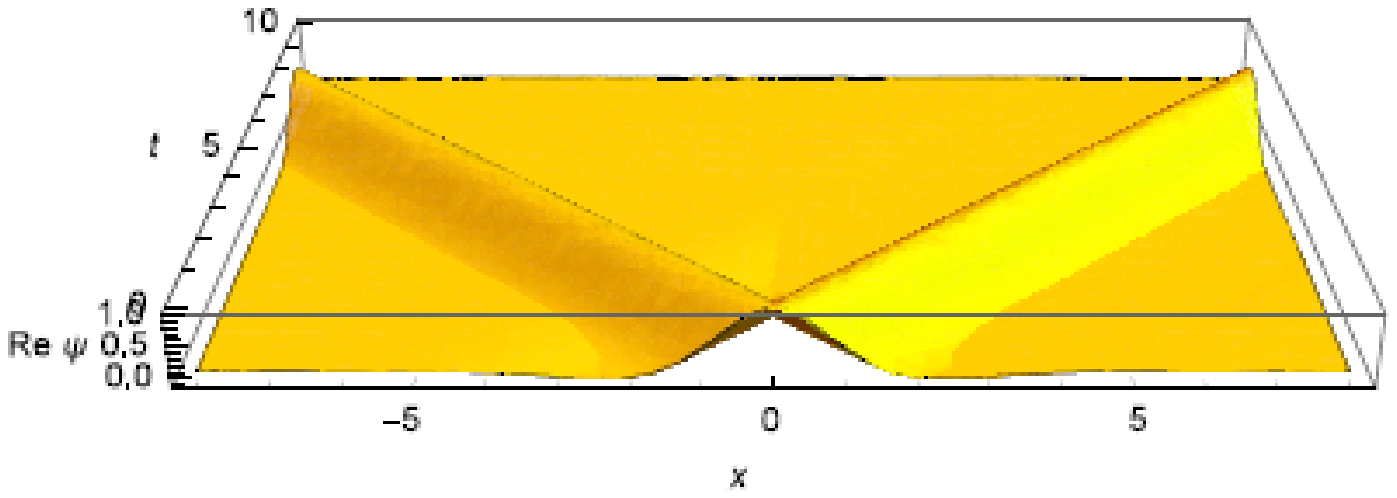} \ \
\includegraphics[width=0.49\textwidth]{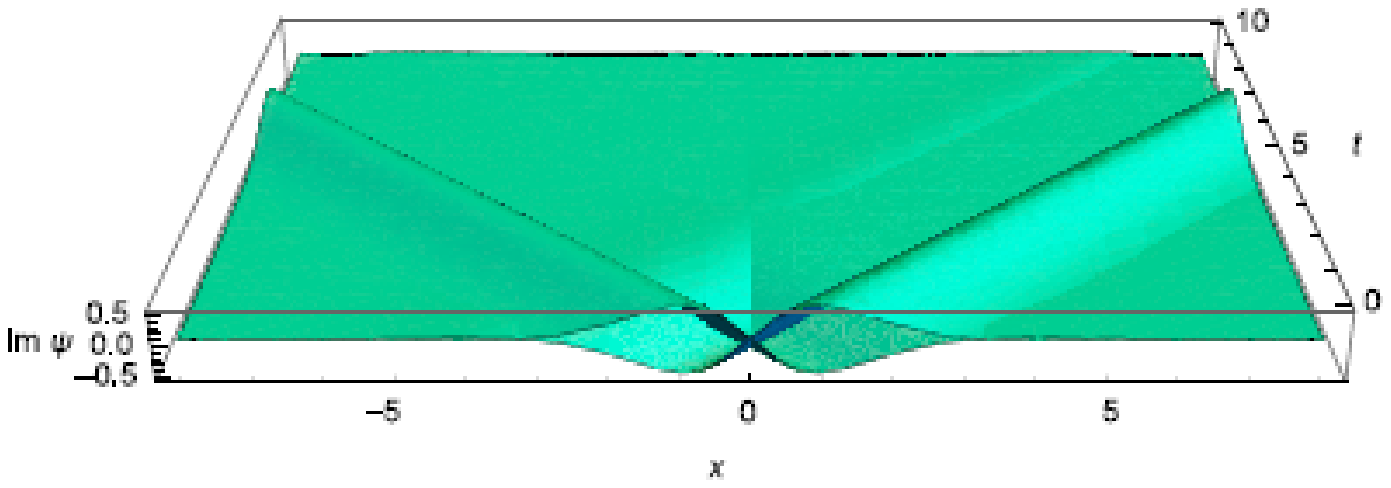}
\includegraphics[width=0.49\textwidth]{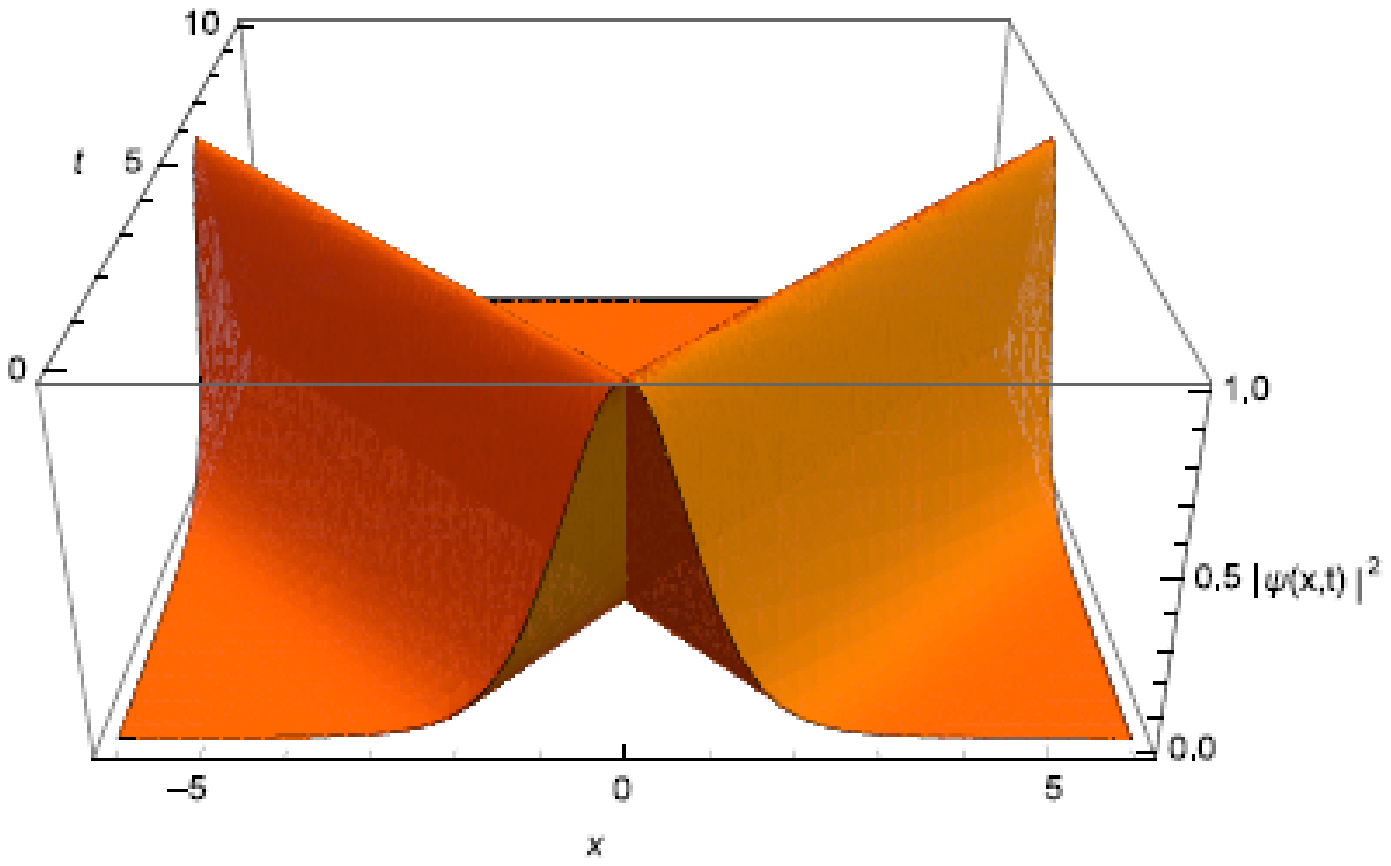}
\caption{\label{ours}
(Color online) Plots of the real and imaginary parts as well as the squared modulus of the traveling soliton solution
(\ref{eq:solucionfin1}) for the case of dimensionless variables and the same parameters as in the previous figure and $V_0=1$.  In this solution, the real and imaginary parts of the wave function have almost fixed shapes during the propagation. 
}\end{center}\end{figure}
In other words, only linearly-extended NLS equations can have traveling soliton solutions of the form
\begin{equation}\label{eq:solucionfin1}
\psi(x,t)=\sqrt{\frac{2\Delta Q}{\gamma}} \,
\mathrm{sech} \! \left[\sqrt{\frac{2m\Delta Q}{\hbar^{2}}}\left(z+z_0\right)\right]
e^{i\left[\frac{mv}{\hbar}z+\phi_0\right]}~.
\end{equation}
This solution, which is plotted in Fig.~(\ref{ours}), has an interesting particular feature that does not occur in the case of the standard solution. It is the solution for a non zero potential which depends exclusively on the traveling variable in both amplitude and phase and in this sense it is also different from the solution (\ref{gauge}) since it cannot be obtained from the standard solution by applying the gauge. Therefore we call (\ref{eq:solucionfin1}) a nongauge bright soliton. Both its amplitude and wave width depend on the {\it excess potential energy $\Delta Q$} with respect to the kinetic energy which keeps into motion a classical particle of mass $m$ at constant speed $v$.

\bigskip


\noindent {\bf 3. One-parameter family of generalized NLS equations}\\

We turn now to the second goal of this work, and show that there exists a one-parameter family of generalized NLS equations with solitonic solutions,
where the parameter of the family is the exponent of the nonlinear term. The NLS equations we consider are linearly extended by a constant potential as previously but the nonlinearity is expressed as an arbitrary function $f(|\psi|)$
\begin{equation}\label{eq:ecnlsg}
i\hbar\frac{\partial\psi}{\partial t}=-\frac{\hbar^{2}}{2m}\frac{\partial^{2}\psi}{\partial x^{2}}-\gamma f(|\psi|)\,\psi+V_0\psi~.
\end{equation}
We pose the same problem of whether the constant potential term allows us to find soliton solutions.
Let us fix the phase as before, $\phi(z)=\frac{mv}{\hbar}z+\phi_0$, and assume that the amplitude has the form
\begin{equation}\label{eq:ka1ka2}
A(z)=c_1\, \mbox{sech}^p(c_2z)~,
\end{equation}
where $c_1$ and $c_2$ are real constants, and $p$ is a strictly positive integer or rational number. Recently, this type of ansatz has been used in an alternative form of the variational approximation for solitons of equations with higher degree of polynomial nonlinearities \cite{m1,m2}. Differentiating twice, one can find that $A(z)$ satisfies
\begin{equation}\label{eq:ampl2n}
\frac{\mathrm{d}^{2}A}{\mathrm{d}z^{2}}-p^{2}c_{2}^{2}A+p(p+1)\left(\frac{c_{2}}{c_{1}^{1/p}}\right)^{2}
A^{\frac{2}{p}}A=0~.
\end{equation}

On the other hand, if we insert the ansatz (\ref{trvw}) together with the linear phase $\phi(z)$, into (\ref{eq:ecnlsg}), we find that the amplitude also satisfies
\begin{equation}\label{eq:gene}
\frac{\mathrm{d}^{2}A}{\mathrm{d}z^{2}}-
\frac{2m{\cal B}}{\hbar^{2}}A+ \frac{2m}{\hbar^{2}}\gamma  f(A) \, A=0~.
\end{equation}

Clearly, if we want to relate equations~(\ref{eq:ampl2n}) and (\ref{eq:gene}), we must have that
\begin{equation}\label{eq:16}
c_{2}=\sqrt{\frac{2m{\cal B}}{p^2\hbar^{2}}}
\end{equation}
and
\begin{equation}\label{eq:17}
f(|\psi|)\equiv f(A)=\frac{p+1}{p}\frac{{\cal B}}{\gamma}\left(\frac{A}{c_1}\right) ^{\frac{2}{p}}~.
\end{equation}
%

Since $p$ is arbitrary, we have constructed an infinite number of generalized NLS equations with soliton solutions. This family includes the NLS equation of constant potential for $p=1$ and other nonlinear equations with soliton solutions for other values of $p$, as we shall show next for the case $p=2$.

\bigskip



\noindent {\bf 4. The $p=2$ generalized NLS equation: Connection with KdV and BBM equations}\\

A lot of work has been developed in order to find connections between the standard NLS equation and other nonlinear differential equations that possess solitonic solutions. That is the case of the KdV \cite{kdv1} and the related BBM equation \cite{BBM}, which are well-established mathematical models of waves on shallow water surfaces and known to be connected to the standard NLS equation only by approximations \cite{Kivshar,Bethuel,Chiron,Boyd}. Here, we show that there exists an exact relationship between these equations and the equation of parameter $p=2$ from the family of generalized NLS equations
\begin{equation}\label{eq:4-1}
\frac{\mathrm{d}^{2}A}{\mathrm{d}z^{2}}-4c_{2}^{2}A+6\left(\frac{c_{2}}{\sqrt{c_{1}}}\right)^{2}A^2=0~.
\end{equation}


For the KdV equation
\begin{equation}
B_t+6BB_x+B_{xxx}=0~,\label{eq:eckdv}
\end{equation}
which, additionally to shallow water waves, is used in electric transmission lines, optical fibers, and other fields \cite{Remoissenet}, we use the traveling wave ansatz (\ref{trvw}) with zero phase and integrate once to find
\begin{equation}
\frac{d^2B}{d z^2}+3B^2-vB=0~,\label{eckdv2}
\end{equation}
where the integration constant has been taken as zero.

\begin{figure}[htpb]
\begin{center}
\includegraphics[width=0.35\textwidth]{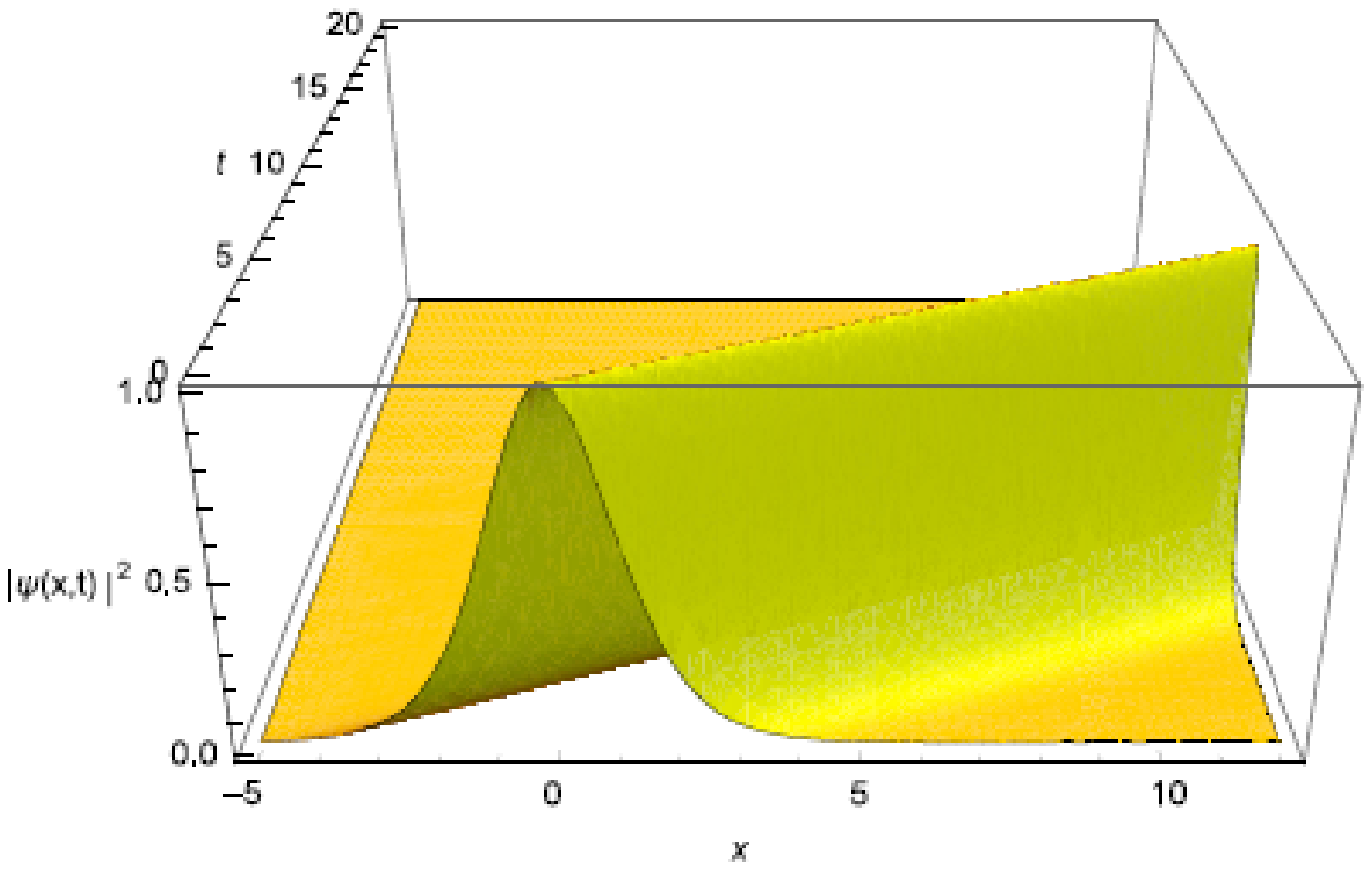} \ \
\includegraphics[width=0.35\textwidth]{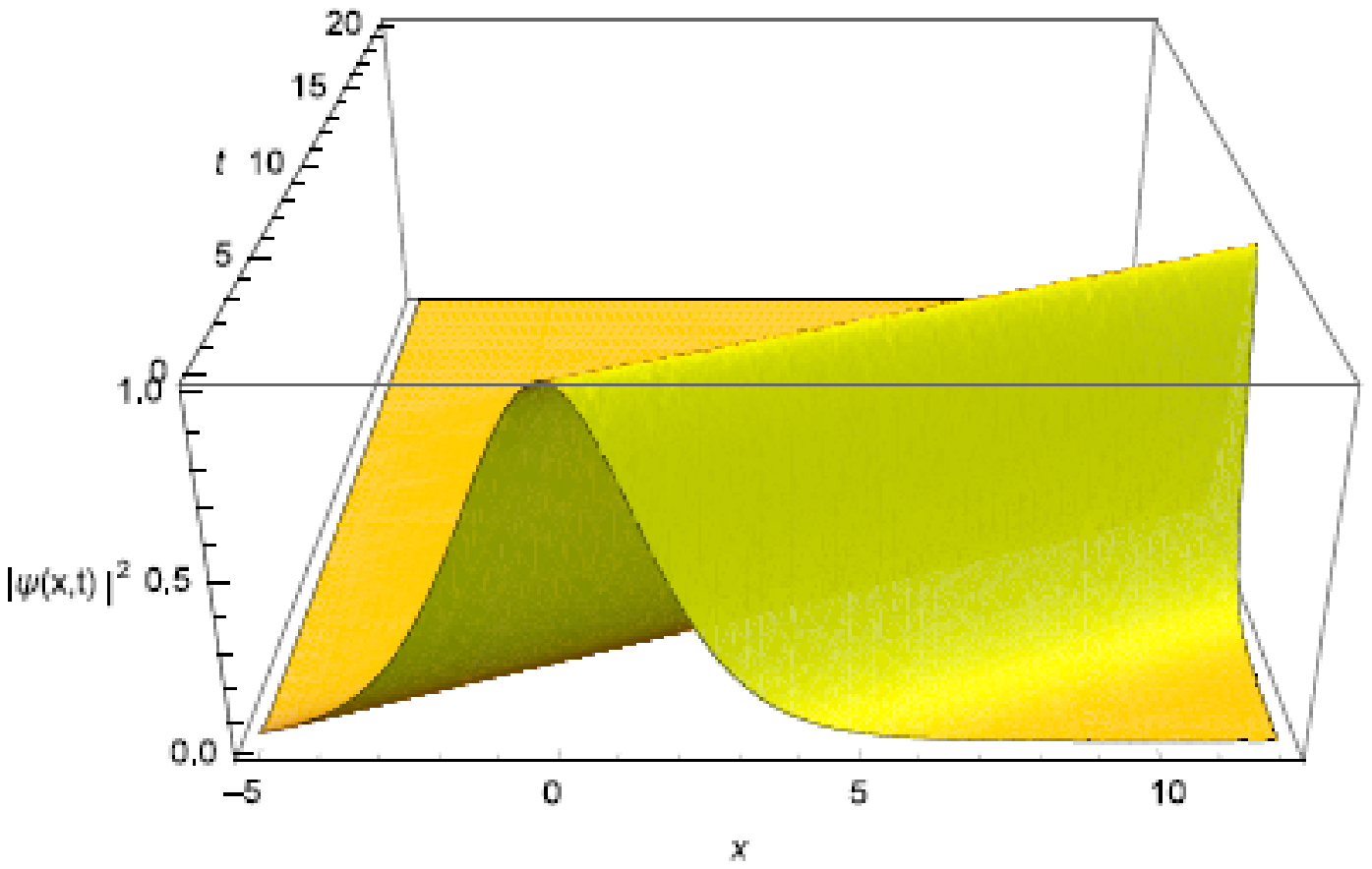}
\caption{\label{waves}
(Color online) Plots of KdV (top) and BBM (bottom) solutions, for $v=2$. The width of the soliton is bigger by $\sqrt{2}$ in the BBM case.
}\end{center}\end{figure}

Assuming $B$ and $B_z$ both going to zero when $z\rightarrow \pm \infty$, one can get the well-known soliton
\begin{equation}
B(x,t)=\frac{v}{2}\mathrm{sech}^{2}\left[\frac{\sqrt v}{2}\left(x-vt\right)\right]~, \label{eq:solkdv}
\end{equation}
which is plotted in Fig.~\ref{waves} (top).

Equations~(\ref{eckdv2}) and~(\ref{eq:4-1}) are the same when
\begin{eqnarray}	
\begin{array}{cccc}
c_1 \equiv \frac{v}{2} \ \ & \ \  c_2\equiv\frac{\sqrt v}{2}\ \  & \ \   \frac{\hbar^{2}}{2m}\equiv 1~,
\end{array}
\label{kdvrel}
\end{eqnarray}
which exactly match the solution (\ref{eq:ka1ka2}) with (\ref{eq:solucionfinal}) if further (\ref{eq:16}) and (\ref{eq:17}) are taken into account.

\medskip

Finally, we refer to the regularized long wave regime of the KdV equation, or the BBM equation. It can be viewed as an alternative to KdV by replacing the dispersion term $B_{xxx}$ with $-B_{xxt}$ to
reflect a bounded dispersion relation, see \cite{BBM,Per}. Thus, we write the BBM equation as follows
\begin{equation}\label{eq:ecbbm}
C_t+6C C_x-C_{xxt}=0~.
\end{equation}

Changing again to the traveling variable, the solution is found from the elliptic equation
\begin{equation}
\frac{d^2C}{d z^2}-\frac{3}{v}C^2+C=0~,\label{ecbbm2}
\end{equation}
where again the integration constant is zero. Then via the identifications
\begin{eqnarray}	
\begin{array}{cccc}
c_1 \equiv \frac{v}{2} \ \ & \ \ c_2\equiv\frac{1}{2}\ \  & \ \  \frac{\hbar^{2}}{2m}\equiv -1
\end{array}
\label{bbmrel}
\end{eqnarray}
in (\ref{eq:ka1ka2}) and (\ref{eq:4-1}), and zero boundary conditions as above, we obtain the known soliton
\begin{equation}
C(x,t)=\frac{v}{2}\mathrm{sech}^{2}\left[\frac{1}{2}\left(x-vt\right)\right]~,\label{eq:solbbm}
\end{equation}
see Fig.~\ref{waves} (bottom).

For complicated cases of the BBM equation modified by viscosity, where the modified elliptic equation (\ref{eckdv2}) is solved via the Weierstrass $\wp$ functions, see \cite{Man} together with the references cited therein.\\

\bigskip

\noindent {\bf 5. Conclusion}\\

We have discussed the `nongauge' bright soliton of a NLS equation extended by a constant potential that cannot occur for the standard NLS equation
(\ref{eq:enls}). A one-parameter family of generalized NLS equations with constant potential having the family parameter determined by the order of nonlinearity is also introduced in this work. Furthermore, it is shown that the KdV and BBM solutions represent the amplitude of the solution of generalized NLS equation with nonlinearity $f(|\psi|)\equiv f(A)$. This points to the possibility of finding relationships between these generalized NLS equations and other nonlinear equations with soliton solutions.

\bigskip

{\bf Acknowledgments}:
The second author wishes to thank the Red de Gravitaci\'on CONACyT for a fellowship.

\end{document}